\documentclass[conference]{IEEEtran}
\usepackage[T1]{fontenc}
\usepackage{txfonts}

\makeatletter
\renewcommand{\section}{\@startsection{section}{1}{\z@}{0.1\baselineskip}%
{0.1\baselineskip}{\normalfont\normalsize\centering\scshape}}%
\renewcommand{\subsection}{\@startsection{subsection}{2}{\z@}{0.1\baselineskip}%
{0.1\baselineskip}{\normalfont\normalsize\itshape}}%
\makeatother

\author{\IEEEauthorblockN{Olav Geil\IEEEauthorrefmark{1},
Ryutaroh Matsumoto\IEEEauthorrefmark{2} and
Diego Ruano\IEEEauthorrefmark{1}}
\IEEEauthorblockA{\IEEEauthorrefmark{1}Department of Mathematical Sciences, Aalborg University, Denmark}
\IEEEauthorblockA{\IEEEauthorrefmark{2}Department of Communications and Integrated Systems, Tokyo Institute of Technology, 152-8550 Japan}
}

\title{List Decoding Algorithms based on Gr\"obner Bases
for General One-Point AG Codes\IEEEauthorrefmark{3}}
\date{April 22, 2012}

\newtheorem{definition}{Definition}
\newtheorem{proposition}[definition]{Proposition}
\newtheorem{theorem}[definition]{Theorem}

\newtheorem{assumption}[definition]{Assumption}

\newcommand{\ev}{\mathrm{ev}}

\newcommand{\wt}{\mathrm{wt}}
\newcommand{\RR}{\mathcal{L}(\infty Q)}
\newcommand{\gw}{\omega}

\begin{document}
\hypersetup{pdfstartview={FitBH -32768},pdfauthor={Olav Geil, Ryutaroh Matsumoto and Diego Ruano},pdftitle={List Decoding Algorithms based on Gr\"obner Bases for General One-Point AG Codes},pdfkeywords={algebraic geometry code, Gr\"obner basis, list decoding}}
\maketitle
\begin{abstract}
We generalize the list decoding algorithm for
Hermitian codes proposed by
Lee and O'Sullivan \cite{lee09} based on Gr\"obner bases
to general one-point AG codes,
under an assumption weaker than one used by Beelen and
Brander \cite{beelen10}.
By using the same principle, we also generalize the
unique decoding algorithm for one-point AG codes
over the Miura-Kamiya $C_{ab}$ curves
proposed by Lee, Bras-Amor\'os and O'Sullivan \cite{lee11}
to general one-point AG codes,
without any assumption.
Finally we extend the latter unique decoding algorithm
to list decoding, modify it so that it can be used with
the Feng-Rao improved code construction,
prove equality between its error correcting capability
and half the minimum distance lower bound by Andersen and Geil
\cite{andersen08} that has not been done
in the original proposal, and remove the unnecessary computational
steps so that it can run faster.
\renewcommand\thefootnote{\IEEEauthorrefmark{3}}
\footnotetext{%
To appear in Proc.\ 2012 IEEE International Symposium on Information
Theory, July 1--6, 2012, Boston, MA, USA.}
\end{abstract}

\section{Introduction}
We consider the list decoding of one-point algebraic geometry (AG) codes.
Guruswami and Sudan \cite{guruswami99} proposed the well-known
list decoding
algorithm for one-point AG codes, which consists of
the interpolation step and the factorization step.
The interpolation step has large computational complexity and
many researchers have proposed faster interpolation steps,
see \cite[Figure 1]{beelen10}.
Lee and O'Sullivan \cite{lee09} proposed a faster interpolation step based
on the Gr\"obner basis theory for one-point Hermitian codes.
Little \cite{little11} generalized their method \cite{lee09}
by using the same assumption as Beelen and Brander \cite[Assumptions 1 and 2]{beelen10}.
The aim of the first part of this paper is to generalize the
method \cite{lee09} to an even wider class of algebraic curves
than \cite{little11}.

The second part proposes another list decoding algorithm
whose error-correcting capability is higher than
\cite{beelen10,guruswami99,lee09,little11} and whose computational
complexity is empirically manageable.
Lee, Bras-Amor\'os and O'Sullivan \cite{lee12,lee11}
proposed a unique decoding (not list decoding)
algorithm for primal codes based on the majority voting
inside Gr\"obner bases.

There were several rooms for improvements in the original
result \cite{lee11}, namely,
(a) they have not clarified
the relation between its error-correcting capability
and existing minimum distance lower bounds 
except for the Hermitian codes, (b) they assumed that the maximum
pole order used for code construction is less than the code length,
and (c) they have not shown how to use the method with the
Feng-Rao improved code construction \cite{feng95}.
In the second part of this paper, we shall
(1) prove that the error-correcting capability of the original
proposal is always equal to half of the bound in
\cite{andersen08} for the minimum distance of one-point
primal codes, (2)
generalize their algorithm to work with any one-point
AG codes, (3) modify their algorithm to a list decoding
algorithm, (4) remove the assumptions (b) and (c) above,
and (5) remove unnecessary computational steps from the
original proposal.
The proposed algorithm is implemented on the Singular computer algebra
system \cite{singular}, and we verified that
the proposed algorithm can correct more errors than 
\cite{beelen10,guruswami99,lee09,little11} with manageable
computational complexity.
The omitted proofs and the implementation of the proposed algorithm
are available as the expanded versions \cite{gmr12a,gmr12b}
of this conference paper.

\section{Notation and Preliminary}\label{sec2}
Our study heavily relies on the standard form of algebraic curves
introduced independently by Pellikaan \cite{geilpellikaan00} and Miura
\cite{miura98}.
Let $F/\mathbf{F}_q$ be an algebraic function field of one variable over
a finite field $\mathbf{F}_q$ with $q$ elements.
Let $g$ be the genus of $F$.
Fix $n+1$ distinct places $Q$, $P_1$, \ldots,
$P_n$ of degree one in $F$ and a nonnegative integer
$u$. We consider the following
one-point algebraic geometry (AG) code
\[
C_u = \{ (f(P_1), \ldots, f(P_n)) \mid f \in \mathcal{L}(uQ)\}.
\]
Suppose that the Weierstrass semigroup $H(Q)$
at $Q$ is generated by
$a_1$, \ldots, $a_t$, and choose
$t$ elements $x_1$, \ldots,
$x_t$ in $F$ whose pole divisors are $(x_i)_\infty = a_iQ$ for $i=1$, \ldots, $t$.
Without loss of generality we may assume the availability of
such $x_1$, \ldots,
$x_t$, because otherwise we cannot find a basis of $C_u$ for every $u$.
Then we have that $\mathcal{L}(\infty Q) = \cup_{i=1}^\infty\mathcal{L}(iQ)$
is equal to $\mathbf{F}_q[x_1$, \ldots, $x_t]$ \cite{saints95}. Let $\mathrm{m}_i$ be the maximal ideal $P_i \cap \mathcal{L}(\infty Q)$ of $\mathcal{L}(\infty Q)$ associated with the place $P_i$. We express $\mathcal{L}(\infty Q)$ as a residue class ring
$\mathbf{F}_q[X_1$, \ldots, $X_t]/I$
of the polynomial ring $\mathbf{F}_q[X_1$, \ldots, $X_t]$, where
$X_1$, \ldots, $X_t$ are transcendental over $\mathbf{F}_q$,
and $I$ is the kernel of the canonical homomorphism sending
$X_i$ to $x_i$. Pellikaan and Miura \cite{geilpellikaan00,miura98}
identified the following convenient representation of
$\mathcal{L}(\infty Q)$ by using the Gr\"obner basis theory \cite{adams94}.
The following review is borrowed from \cite{miuraform}.
Hereafter, we assume that the reader is familiar with the
Gr\"obner basis theory in \cite{adams94}.

Let $\mathbf{N}_0$ be the set of nonnegative integers.
For $(m_1$, \ldots, $m_t)$, $(n_1$, \ldots, $n_t) \in
\mathbf{N}_0^t$,
we define the monomial order $\succ$
such that $(m_1$, \ldots, $m_t)$ $\succ$ $(n_1$, \ldots, $n_t)$
if $a_1 m_1 + \cdots + a_t m_t > a_1 n_1 + \cdots + a_t n_t$,
or $a_1 m_1 + \cdots + a_t m_t = a_1 n_1 + \cdots + a_t n_t$,
and $m_1 = n_1$, $m_2 = n_2$, \ldots,
$m_{i-1} = n_{i-1}$, $m_i<n_i$, for some $1 \leq i \leq t$.
Note that a Gr\"obner basis of $I$ with respect to $\succ$
can be computed by \cite[Theorem 15]{saints95} or
\cite[Theorem 4.1]{tang98},
starting from any affine defining equations of $F/\mathbf{F}_q$.

For $i=0$, \ldots, $a_1-1$,
we define $b_i = \min\{ m \in H(Q) \mid m \equiv i \pmod{a_1}\}$,
and $L_i$ to be the minimum element $(m_1$, \ldots,
$m_t) \in \mathbf{N}_0^t$
with respect to $\prec$ such that $a_1 m_1 + \cdots +
a_t m_t = b_i$.
Then we have $\ell_1 = 0$ if we write
$L_i$ as $(\ell_1$, \ldots, $\ell_t)$.
For each $L_i = (0$, $\ell_{i2}$, \ldots, $\ell_{it})$,
define $y_i = x_2^{\ell_{i2}} \cdots x_t^{\ell_{it}} \in \mathcal{L}(\infty Q)$.

The footprint of $I$, denoted by $\Delta(I)$,
is $\{ (m_1$, \ldots, $m_t) \in \mathbf{N}_0^t \mid X_1^{m_1} \cdots
X_t^{m_t}$ is not the leading monomial of
any nonzero polynomial in $I$ with respect to $\prec\}$,
and define $B = \{x_1^{m_1} \cdots x_t^{m_t} \mid 
(m_1$, \ldots, $m_t) \in \Delta(I)\}$.
Then $B$ is a basis of $\mathcal{L}(\infty Q)$
as an  $\mathbf{F}_q$-linear space \cite{adams94},
two distinct elements in $B$ have different pole orders at $Q$,
and
\begin{eqnarray}
&& B \nonumber\\
&=& \{ x_1^m x_2^{\ell_2} \cdots, x_t^{\ell_t} \mid m \in \mathbf{N}_0,
(0, \ell_2, \ldots, \ell_t) \in \{L_0, \ldots, L_{a_1-1}\}\}\nonumber\\
&=& \{ x_1^m y_i \mid m \in \mathbf{N}_0, i=0, \ldots, a_1-1\}.
\label{eq:footprintform}
\end{eqnarray}
Equation (\ref{eq:footprintform}) shows that
$\mathcal{L}(\infty Q)$ is a free $\mathbf{F}_q[x_1]$-module
with a basis $\{y_0$, \ldots, $y_{a_1-1}\}$.

Let $v_Q$ be the unique valuation in $F$ associated with the place $Q$. The semigroup $S=H(Q)$ is equal to $S=\{i a_1 - v_Q (y_j) \mid 0\le i,0\le j<a_1\}$. For each nongap $s\in S$ there is a unique monomial $x_1^i y_j \in \RR$ with $0\le j<a_1$ such that $-v_Q (x_1^i  y_j )=s$ by \cite[Proposition 3.18]{miuraform},
and let us denote this monomial by $\varphi_s$. 
Let $\Gamma \subset S$, and we may consider the one-point codes
\begin{equation}
C_\Gamma = \langle ( \varphi_s (P_1), \ldots ,  \varphi_s (P_n) ) \mid s \in \Gamma \rangle.\label{eq:cgamma}
\end{equation}
One motivation for considering these codes is
that it was shown in \cite{andersen08} how to increase the dimension of the one-point codes without decreasing the bound for the minimum distance.

\section{Generalization of Lee-O'Sullivan's List Decoding to General One-Point AG Codes}\label{sec3}
\subsection{Background on Lee-O'Sullivan's Algorithm}
In the famous list decoding algorithm for the one-point AG
codes in \cite{guruswami99},
we have to compute the univariate interpolation polynomial
whose coefficients belong to $\mathcal{L}(\infty Q)$.
Lee and O'Sullivan \cite{lee09}
proposed a faster algorithm to compute
the interpolation polynomial for the Hermitian one-point codes.
Their algorithm was sped up and generalized to one-point AG codes
over the so-called $C_{ab}$ curves \cite{miura92} by Beelen and
Brander \cite{beelen10} with an additional assumption.
In this section we generalize Lee-O'Sullivan's procedure to
general one-point AG codes with an  assumption weaker than \cite[Assumption 2]{beelen10},
which will be introduced in and used after Assumption \ref{assump1}.

Let $m$ be the multiplicity parameter in \cite{guruswami99}.
Lee and O'Sullivan introduced the ideal $I_{\vec{r},m}$
containing the interpolation polynomial corresponding to
the received word $\vec{r}$ and the multiplicity $m$.
The ideal $I_{\vec{r},m}$ contains the interpolation polynomial
as its minimal nonzero element with respect to the monomial order. 
We will give another module $I'_{\vec{r},m}$ for general
algebraic curves, from which we can also obtain the required
interpolation polynomial.

\subsection{Definition of the Interpolation Ideal}
Let $\vec{r} = (r_1$, \ldots, $r_n) \in \mathbf{F}_q^n$
be the  received word.
For a divisor $G$ of $F$, we define
$\mathcal{L}(-G + \infty Q) = \bigcup_{i=1}^\infty \mathcal{L}(-G+iQ)$.
We see that $\mathcal{L}(-G + \infty Q)$ is an ideal of $\mathcal{L}(\infty Q)$
\cite{matsumoto99ldpaper}.

Let $h_{\vec{r}} \in \mathcal{L}(\infty Q)$ such that $h_{\vec{r}}(P_i) = r_i$.
Computation of such $h_{\vec{r}}$ is easy provided that
we can construct generator matrices for $C_u$
for every $u$. We can choose $h_{\vec{r}}$ so that $-v_Q(h_{\vec{r}}) \leq n+ 2g-1$.

Let $Z$ be transcendental over $\mathcal{L}(\infty Q)$,
and $D=P_1 + \cdots + P_n$.
Define the set $
I'_{\vec{r},m} = \{ Q(Z) \in \mathcal{L}(\infty Q)[Z] \mid Q(Z)$ has multiplicity $m$ for
all $(P_i$, $r_i) \}$.
This definition of the multiplicity is the same as \cite{guruswami99}.
Therefore, we can find the interpolation polynomial used in \cite{guruswami99}
from $I'_{\vec{r},m}$.
We shall explain how to find efficiently the interpolation polynomial
from $I'_{\vec{r},m}$.

For $i=0$, \ldots, $m$ and $j=0$, \ldots, $a_1-1$,
let $\eta_{i,j}$ to be an element in $\mathcal{L}(-iD+\infty Q)$
such that $-v_Q(\eta_{i,j})$ is the minimum among $\{
-v_Q(\eta) \in \mathcal{L}(-iD+\infty Q) \mid -v_Q(\eta) \equiv j \pmod{a_1}\}$.
Such elements $\eta_{i,j}$ can be computed by \cite{matsumoto99ldpaper}
before receiving $\vec{r}$.
It was also shown \cite{matsumoto99ldpaper} that
$\{ \eta_{i,j} \mid j=0$, \ldots, $a_1-1 \}$ generates
$\mathcal{L}(-iD+\infty Q)$ as an $\mathbf{F}_q[x_1]$-module.
Note also that we can choose $\eta_{0,i} = y_i$ defined in Sec.\ \ref{sec2}.
By Eq.\ (\ref{eq:footprintform}),
all $\eta_{i,j}$
and $h_{\vec{r}}$ can be
expressed as polynomials in $x_1$ and $y_0$, \ldots, $y_{a_1-1}$.
Thus we have
\begin{theorem}\label{thm:generators}
Let $\ell \geq m$.
$\{ (Z-h_{\vec{r}})^{m-i}\eta_{i,j} \mid i=0$, \ldots, $m$, $j=0$, \ldots, $a_1-1 \}$
$\cup \{ Z^{\ell-m}(Z-h_{\vec{r}})^{m}\eta_{0,j} \mid \ell=1$, \ldots, $j=0$, \ldots, $a_1-1 \}$
generates $I_{\vec{r},m,\ell} = I_{\vec{r},m} \cap \{ Q(Z) \in \mathcal{L}(\infty Q)[Z] \mid \deg_Z Q(Z) \leq \ell \}$ as an $\mathbf{F}_q[x_1]$-module.
\end{theorem}

\subsection{Computation of the Interpolated Polynomial from the
Interpolation Ideal $I_{\vec{r},m}$}\label{sec32}
For $(m_1$, \ldots, $m_t$, $m_{t+1})$, $(n_1$, \ldots, $n_t$, $n_{t+1}) \in
\mathbf{N}_0^{t+1}$,
we define the monomial order $\succ_u$
in $\mathbf{F}_q[X_1$, \ldots, $X_t$, $Z]$
such that $(m_1$, \ldots, $m_t$, $m_{t+1})$ $\succ$ $(n_1$, \ldots, $n_t$, $n_{t+1})$
if $a_1 m_1 + \cdots + a_t m_t + u m_{t+1}> a_1 n_1 + \cdots + a_t n_t + un_{t+1}$,
or $a_1 m_1 + \cdots + a_t m_t + u m_{t+1} = a_1 n_1 + \cdots + a_t n_t + u n_{t+1}$,
and $m_1 = n_1$, $m_2 = n_2$, \ldots,
$m_{i-1} = n_{i-1}$, $m_i<n_i$, for some $1 \leq i \leq t+1$.
As done in \cite{lee09},
the interpolation polynomial is the smallest nonzero polynomial
with respect to $\succ_u$ in the preimage of $I_{\vec{r},m}$.
Such a smallest element can be found from a Gr\"obner basis of the
$\mathbf{F}_q[x_1]$-module $I_{\vec{r},m,\ell}$
in Theorem \ref{thm:generators}.
To find such a Gr\"obner basis, Lee and O'Sullivan proposed the following
general purpose algorithm as \cite[Algorithm G]{lee09}.

Their algorithm \cite[Algorithm G]{lee09}
efficiently finds a Gr\"obner basis of submodules of $\mathbf{F}_q[x_1]^s$
for a special kind of generating set and monomial orders.
Please refer to \cite{adams94} for Gr\"obner bases for modules.
Let $\mathbf{e}_1$, \ldots, $\mathbf{e}_s$ be the standard basis
of $\mathbf{F}_q[x_1]^s$.
Let $u_x$, $u_1$, \ldots, $u_s$ be positive integers.
Define the monomial order in $\mathbf{F}_q[x_1]^s$
such that $x_1^{n_1} \mathbf{e}_i \succ_{\mathrm{LO}} x_1^{n_2} \mathbf{e}_j$
if ${n_1}u_x + u_i > {n_2}u_x + u_j$ or ${n_1}u_x + u_i = {n_2}u_x + u_j$
and $i>j$.
For $f = \sum_{i=1}^s f_i(x_1)\mathbf{e}_i \in \mathbf{F}_q[x_1]^s$,
define $\mathrm{ind}(f) = \max\{i \mid f_i(x_1) \neq 0\}$,
where $f_i(x_1)$ denotes a univariate polynomial in $x_1$ over $\mathbf{F}_q$.
Their algorithm \cite[Algorithm G]{lee09}
efficiently computes a Gr\"obner basis of a module generated
by $g_1$, \ldots, $g_s \in \mathbf{F}_q[x_1]^s$ such that
$\mathrm{ind}(g_i) = i$.
The computational complexity is also evaluated in \cite[Proposition 16]{lee09}.

Let $\ell$ be the maximum $Z$-degree of the interpolation polynomial
in \cite{guruswami99}. 
The set $I_{\vec{r},m,\ell}$ in Theorem \ref{thm:generators}
is an $\mathbf{F}_q[x_1]$-submodule of $\mathbf{F}_q[x_1]^{a_1(\ell+1)}$
with the module basis $\{ y_j Z^k \mid j=0$, \ldots, $a_1-1$, $k=0$, \ldots,
$\ell \}$.

\begin{assumption}\label{assump1}
We assume that we have $f \in \mathcal{L}(\infty Q)$ whose
zero divisor $(f)_0 = D$.
\end{assumption}
Observe that Assumption \ref{assump1} is implied by \cite[Assumption 2]{beelen10} 
and is weaker than \cite[Assumption 2]{beelen10}.
Let $\langle f \rangle$ be the ideal of $\mathcal{L}(\infty Q)$
generated by $f$. By \cite[Corollary 2.3]{matsumoto99ldpaper}
we have $\mathcal{L}(-D + \infty Q) = \langle f \rangle$.
By \cite[Corollary 2.5]{matsumoto99ldpaper}
we have $\mathcal{L}(-iD + \infty Q) = \langle f^i \rangle$.

Without loss of generality we may assume existence of
$x' \in \mathcal{L}(\infty Q)$ such that $f \in \mathbf{F}_q[x']$.
By changing the choice of $x_1$, \ldots, $x_t$ if necessary,
we may assume $x_1 = x'$ and $f \in \mathbf{F}_q[x_1]$ without
loss of generality, while it is better to make $-v_Q(x_1)$ as small
as possible in order to reduce the computational complexity.
Under the assumption $f \in \mathbf{F}_q[x_1]$,
$f^i y_j$ satisfies the required condition for
$\eta_{i,j}$ in Theorem \ref{thm:generators}.
By naming $y_j z^k$ as $\mathbf{e}_{1+j+ku}$,
the generators in Theorem \ref{thm:generators} satisfy the
assumption in \cite[Algorithm G]{lee09} and we can efficiently compute
the interpolation polynomial required in the list decoding algorithm
in \cite{guruswami99}.

\begin{proposition}
We assign the weight $-iv_Q(x_1)-v_Q(y_j)+ku$ to the module element
$x_1^iy_jz^k$ when we use \cite[Algorithm G]{lee09} to find
the minimal Gr\"obner basis of $I_{\vec{r},m,\ell}$.
Under Assumption \ref{assump1},
the number of multiplications in \cite[Algorithm G]{lee09}
with the generators in Theorem \ref{thm:generators} 
is at most 
\begin{equation}
[\max_j\{-v_Q(y_j)\} + m (n+2g-1)+ u(\ell-m)]^2a_1^{-1}
\sum_{i=1}^{a_1(\ell + 1)} i^2. \label{eq:compl}
\end{equation}
\end{proposition}

\section{New List Decoding based on Majority Voting inside Gr\"obner Bases}\label{sec4}

A unique decoding algorithm for one-point codes over $C_{ab}$ curves has recently been  introduced in \cite{lee11}. This algorithm is also based on the interpolation approach, an ideal containing the interpolation polynomials of a received word is computed. Moreover, the algorithm in \cite{lee11} combines the interpolation approach with syndrome decoding with majority voting scheme. However, this algorithm only considers the non-improved code $C_u$ assuming that $u < n$.

The aim of this section is to extend this algorithm for one-point codes defined over general curves without assuming $u < n$, besides, the modified algorithm performs list decoding. Furthermore, we can speed up the algorithm and deal with Feng-Rao improved codes by changing the majority voting. Still, the main structure of the algorithm remains the same. We stress that we do not assume Assumption \ref{assump1} in this section.

Let $F/\mathbf{F}_q$ be an algebraic function field as in Sec.\  \ref{sec2}, we consider the same notation and concepts already introduced in Secs.\ \ref{sec2} and \ref{sec3}. Let $\Gamma= \{s_1, s_2 , \ldots , s_k \} \subset S$ and consider the code $C_\Gamma$ defined in Eq.\ (\ref{eq:cgamma}). We will assume that $\Gamma = \Gamma_{\mathrm{indep}}$, where\begin{equation}\label{eq:improved}\Gamma_{\mathrm{indep}} = \{ s \in \Gamma \mid \ev(\varphi_s) \notin \langle \ev(\varphi_{s'}) : s' \in \Gamma, s' < s \rangle \},\end{equation}since there is no interest in considering  $s \in \Gamma \setminus  \Gamma_{\mathrm{indep}}$. Let $\vec{r}$ be a received word.
Choose \textbf{any codeword in}
$C_\Gamma$ as $\vec{c}$ and define $\vec{e}(\vec{c})= \vec{r} - \vec{c}$.
Then there is a unique
\begin{equation}
\mu = \sum_{s \in \Gamma} \omega_s \varphi_s,
\label{eq:mus}
\end{equation}
with $\vec{c} = \ev(\mu) = (\mu (P_1), \ldots , \mu (P_n))$.

As in  Sec.\  \ref{sec32}, we consider $\RR$  as an $\mathbf{F}[x_1]$-module of rank $a_1$ with basis $\{y_j \mid 0 \le j < a_1 \}$. For $f \in \mathbf{F} [x_1]$, we denote by $f[x_1^k]$ the coefficient of the term $x_1^k$ in $f$.

The following ideal containing the interpolation polynomial for a received word $\vec{r}$ is defined in \cite{lee11},
\[
I_{\vec{r}} = \{f(z) \in  \mathcal{L}(\infty Q)z \oplus \mathcal{L}(\infty Q) \mid v_{P_i}(f(r_i)) \geq 1,~1\le i \le n\}.
\]

Moreover, $I_{\vec{r}}$ is a special case of the interpolation ideal in \cite{lee09}. Thus, by Sec.\  \ref{sec3}, we have that $\mathcal{L}(\infty Q) z \oplus \mathcal{L}(\infty Q)$ is a free $\mathbf{F}_q [x_1]$-module of rank $2a_1$ with basis  $\{y_j z , y_j \mid 0 \le j < a_1 \}$.  Hence an element in $\mathcal{L}(\infty Q) z \oplus \mathcal{L}(\infty Q)$ can be uniquely expressed by monomials in$$\Omega_1 = \{x_1^i y_j z^k\mid 0\le i, 0\le j<a_1, 0\le k \le 1\}.$$
Recall also that an element in $\mathcal{L}(\infty Q)$ can be uniquely expressed by monomials in $\Omega_0 = \{x_1^i y_j\mid 0\le i, 0\le j<a_1\}.$

By the previous section,$$G=\{\eta_0,\eta_1,\ldots,\eta_{a_1-1},z-h_{\vec{r}},y_1 (z-h_{\vec{r}}),\dots, y_{a_1-1} (z-h_{\vec{r}})\},$$with $\eta_i$ and $h_{\vec{r}}$ as in Sec.\  \ref{sec3}, is a Gr\"obner basis of the $\mathbf{F}_q[x_1]$-module
$I_{\vec{r}}$ with respect to the monomial order $>_{-v_Q(h_{\vec{r}})}$ defined in Sec.\  \ref{sec32}.  

Let $J_{\vec{e}(\vec{c})}=\cap_{e_i\neq 0}\mathrm{m}_i$ be the ideal of the error vector and let $\epsilon_i \in \mathcal{L}(\infty Q)$ such that $-v_Q (\epsilon_i)$ is the minimum among $\{ f \in J_{\vec{e}(\vec{c})} \mid -v_Q (f) \equiv i \pmod{a_1}  \}$, for $i=0, \ldots , a_1 -1$. One has that $\{\epsilon_0,\epsilon_1,\dots,\epsilon_{a_1-1}\}$ is a module-Gr\"obner basis with respect to the restriction to $\mathcal{L}(\infty Q)$ of the order $>_u$ introduced in Sec.\  \ref{sec32} (which is independent of $u$). Note that  $-v_Q(J_{\vec{e}(\vec{c})}) =  \{ s -v_Q(\epsilon_i) | 0 \le i<a_1, s \in S \}$. Then
\begin{equation} 
	\sum_{0\le i<a_1}\deg_{x_1}(\mathrm{LT}(\epsilon_i))=\dim_{\mathbf{F}} \RR/J_{\vec{e}(\vec{c})}=\wt(\vec{e}(\vec{c})).
\end{equation}

Before describing the algorithm, we remark that its correctness  is based in a straightforward generalization of some results in \cite[Sec.\  III-A]{lee11}. In particular, we will directly refer  to these results in the description of the algorithm, because the same proofs in \cite{lee11} will hold after considering $y_j$ instead of $y^j$ and $\mathrm{prec} (s)$ instead of $s-1$, where  $\mathrm{prec} (s)=\max \{s' \in S : s' < s\}$, for $s\in S$.  The reader should also be aware that  in this section we follow the notation of previous sections, however, the notation in \cite{lee11} is different. Namely, $P_{\infty}$ denotes $Q$, $R$ denotes $\RR$, $\delta$ denotes $-v_Q$, $x$ denotes $x_1$ and the semigroup $S$ is  the one generated by  $\{  a ,  a_1 , \ldots ,  a_t \}$ in \cite{lee11}.

\subsection{Decoding Algorithm}\label{subsec:Algo}

We can now describe the extension of the algorithm in \cite{lee11}.  For a constant $\tau \in \mathbf{N}$ the following procedure finds all the codewords within Hamming distance $\tau$ from the received word $\vec{r}$

\begin{enumerate}

\item \textit{Initialization:} Let $N = -v_Q (h_{\vec{r}})$ and $G$ be the Gr\"obner basis of the $\mathbf{F}_q[x_1]$-module $I_{\vec{r}}$ defined above. Let $\vec{r}^{(s_k)} =\vec{r}$ and $B^{(s_k)} = G$.  We consider now the steps \textit{Pairing, Voting, Rebasing} for $s \in S \cap [0,N]$ in decreasing order until the earlier termination condition is verified or, otherwise, until $s=s_1$. 

\item \textit{Pairing:}  We consider that
\begin{equation}\vec{r}^{(s)}= \vec{e'}+\ev(\mu^{(s)}),\ \mu^{(s)}=\omega'_s\varphi_s+\mu^{({ \mathrm{prec}(s)})},\ \mu^{({ \mathrm{prec}(s)})}\in L_{ { \mathrm{prec}(s)}}
\label{eq:errorprime}
\end{equation}
and we will determine $\omega'_s$ by majority voting in step 3) provided that $\wt(\vec{e'}) \leq \tau$. Let $B^{(s)}= \{g_i^{(s)},f_i^{(s)}\mid 0\le i<a_1\}$ be a Gr\"obner basis of the $\mathbf{F}_q[x_1]$-module $I_{\vec{r}^{(s)}}$ with respect to $>_s$ where 
$$
	 g_i^{(s)} = \sum_{0\le j<a_1}c_{i,j}  y_j  z+ \sum_{0\le j<a_1} d_{i,j}    y_j, \mathrm{~with~} c_{i,j}, d_{i,j} \in\mathbf{F}_q[x_1],$$	$$ f_i^{(s)}   =   \sum_{0\le j<a_1}a_{i,j}   y_j  z+ \sum_{0\le j<a_1}b_{i,j}   y_j, \mathrm{~with~} a_{i,j}, b_{i,j} \in\mathbf{F}_q[x_1],
$$
and let $\nu_i^{(s)}=\mathrm{LC} (d_{i,i})$.
We assume that $\mathrm{LT}(f_i^{(s)}) = a_{i,i}   y_i  z$ and
$\mathrm{LT}(g_i^{(s)}) = d_{i,i}   y_i$.
By \cite[Lemmas 2,3,4]{lee11}, one has that $$\sum_{0\le i<a_1}\deg(a_{i,i})+\sum_{0\le i<a_1}\deg(d_{i,i})=n,$$and $-v_Q (a_{i,i} y_i) \le -v_Q (\epsilon_i)$ and  $-v_Q (d_{i,i} y_i) \le -v_Q (\eta_i)$ or, equivalently, $\deg (a_{i,i} ) \le \deg_{x_1} (\mathrm{LT} (\epsilon_i))$ and $\deg (d_{i,i} ) \le \deg_{x_1} (\mathrm{LT} (\eta_i))$, for $0 \le i < a_1$.

For $0 \le i <a_1$, there are unique integers $0\le i'<  a_1$ and $k_i$ satisfying$$-v_Q(a_{i,i} y_i) + s = a_1 k_i -v_Q(y_{i'}).$$Note that by the definition above\begin{equation}\label{eq:iprime}i' = i + s \bmod  a_1,\end{equation}and the integer $-v_Q (a_{i,i} y_i )+s$ is a nongap if and only if $k_i\ge 0$. Now let $c_i=\deg_x(d_{i',i'})-k_i$. Note that the map $i\mapsto i'$ is a permutation of $\{0,1,\dots,a-1\}$ and that  the integer $c_i$ is defined such that $a_1 c_i= -v_Q (d_{i',i'}  y_{i'} )+v_Q (a_{i,i}  y_i )-s$.

\item \textit{Voting:} For each $i \in  \{0, \ldots a_1 -1\}$, we set $$\mu_i= \mathrm{LC}( a_{i,i} y_i \varphi_s), ~ w_i=-\frac{b_{i,i'}[x^{k_i}]}{\mu_i}, ~  \bar{c}_i=\max\{c_i,0\}.$$We remark that the leading coefficient $\mu_i$ must be considered after expressing $a_{i,i} y_i \varphi_s$ by monomials in $\Omega_0$.

Let \begin{equation}\label{eq:nu}\nu(s)=\frac{1}{a_1}\sum_{0\le i<a_1}\max\{-v_Q(\eta_{i'})+v_Q(  y_i )-s,0\}.\end{equation} The error correction capability of the algorithm will be determined by the values $\nu(s)$. The number $\nu(s)$ was introduced in \cite[Proposition 10]{lee11}, we will show in Proposition \ref{prop:AG} that it is equivalent to the cardinality of some sets introduced in \cite{andersen08} for bounding the minimum distance. 

We consider two different candidates depending on whether $s \in \Gamma$ or not:
\begin{itemize}
\item If $s \in S \setminus \Gamma$, set $w=0$. 
\item If $s \in \Gamma$, let $w$ be the element of $\mathbf{F}_q$ with
\begin{equation}\label{eq:acceptedvote}
	\sum_{w=w_i}\bar{c}_i \geq \sum_{w\neq w_i}\bar{c}_i - 2\tau+\nu(s),
\end{equation}since by Proposition \ref{prop10} we will have that $$\sum_{w_i=\gw'_s}\bar{c}_i\geq \sum_{w_i\neq\gw'_s}\bar{c}_i - 2\wt(\vec{e'})+\nu(s),$$
where $\gw'_s$ and $\vec{e'}$ are as defined at Eq.\ (\ref{eq:errorprime}).
\end{itemize}

Let $w_s=w$. If several $w$'s satisfy the condition above, repeat the rest of the algorithm for each of them.  As $s$ decreases, $\nu(s)$ increases and at some point we have $2\tau<\nu(s)$ and at that point at most one $w$ verifies condition (\ref{eq:acceptedvote}).

An interesting difference to the Feng-Rao majority voting is as follows: In the Feng-Rao voting, when $\wt(\vec{e})$ is large, voting for the correct codeword can disappear, i.e., there can be no vote for the correct codeword. In contrast to this, in the Gr\"obner based majority voting, the correct codeword always has a vote, because $I_{\vec{r}}$ contains all the possible codewords and errors.
 
\item \textit{Rebasing:} We consider the automorphism of $\RR [z]$ given by $z \mapsto z + w \varphi_s$ that preserves the leading terms with respect to $>_s$. Hence $B^{(s)}$ is mapped to a set which is  a Gr\"obner basis of $\{f(z + w \varphi_s) \mid f \in I_{\vec{r}^{(s)}} \}$ with respect to $>_s$. However, this set is not (in general) a Gr\"obner basis with respect to $>_{\mathrm{prec}(s)}$, which will be used in the next iteration. Thus, we will update it, for each $i \in \{0, \ldots a_1 -1\}$:

\begin{itemize}
\item If $w_i=w$, then let
\begin{eqnarray*}\label{fkfmvd}
	&g_{i'}^{( \mathrm{prec}(s))}=g_{i'}^{(s)}(z+w\varphi_s),\\
	&f_i^{( \mathrm{prec}(s))}=f_i^{(s)}(z+w\varphi_s),
\end{eqnarray*}where the parentheses denote substitution of the variable $z$ and let $\nu_{i'}^{(  \mathrm{prec}(s))}=\nu_{i'}^{(s)}$.

\item If $w_i\neq w$ and $c_i>0$, then let
$$\begin{array}{ll}
	  g_{i'}^{( \mathrm{prec}(s))}=f_i^{(s)}(z+w\varphi_s)  \\
	 f_i^{( \mathrm{prec}(s))}=x^{c_i}f_i^{(s)}(z+w\varphi_s)-\frac{\mu_i(w-w_i) }{\nu_{i'}^{(s)}}g_{i'}^{(s)}(z+w\varphi_s)
\end{array}$$
and let $\nu_{i'}^{( \mathrm{prec}(s))}=\mu_i(w-w_i)$.
\item If $w_i\neq w$ and $c_i\le 0$, then let
$$\begin{array}{ll}
	 g_{i'}^{({ \mathrm{prec}(s)})}=g_{i'}^{(s)}(z+w\varphi_s)\\
 f_i^{({ \mathrm{prec}(s)})}=f_i^{(s)}(z+w\varphi_s)-\frac{\mu_i(w-w_i)}{\nu_{i'}^{(s)}}x^{-c_i}g_{i'}^{(s)}(z+w\varphi_s)
\end{array}$$
and let $\nu_{i'}^{( \mathrm{prec}(s))}=\nu_{i'}^{(s)}$. \end{itemize}By \cite[proposition 5]{lee11} we have that $$B^{({\mathrm{prec}(s)})}=\{g_i^{({ \mathrm{prec}(s)})}, ~f_i^{({ \mathrm{prec}(s)})}  \mid 0\le i<a_1\},$$  is a Gr\"obner basis of 
$\{f(z + w \varphi_s) \mid f \in I_{\vec{r}^{(s)}} \}
  = I_{\vec{r}^{(\mathrm{prec}(s))}}$ with respect to $>_{\mathrm{prec}(s)}$,
where $\vec{r}^{(\mathrm{prec}(s))} =
\vec{r}^{(s)} - \ev(w \varphi_s)$. We remark that the new Gr\"obner basis $B^{({\mathrm{prec}(s)})}
$ must be considered after expressing it by monomials in $\Omega_1$.

\item \textit{Earlier termination:} The module $I_{\vec{r}}$ is a curve theoretic generalization of
the genus zero case considered in \cite[Definition 9]{kuijper11}. Let $f_{\mathrm{min}} = \alpha_0 + z \alpha_1$ having the smallest $-v_Q(\alpha_1)$ among $f^{(\mathrm{prec}(s))}_0$, \ldots, $f^{(\mathrm{prec}(s))}_{a_1-1}$. When the genus is zero and the number of errors is less than half the minimum distance, we can immediately find the codeword by $-\alpha_0/\alpha_1$
\cite[Theorem 12]{kuijper11}.

Besides, as $s$ decreases, the code $C_{\Gamma^{(s)}}$ treated by each iteration in this algorithm shrinks, where $\Gamma^{(s)} = \{ s' \in \Gamma \mid s' \leq s \}$, while the number of errors remains the same, at some point its minimum distance becomes relatively large compared to the number of errors. Then  $f_{\mathrm{min}}$ should provide the codeword by $-\alpha_0/\alpha_1$. Actually, this phenomenon has also been verified by our computer experiments in Sec.\  \ref{sec34}.

Hence, we propose the following earlier termination criterion: Let $d_{\mathrm{AG}}(C_{\Gamma}) = \min_{s\in \Gamma}\nu(s)$ be the bound for the minimum distance in \cite{andersen08}.  If
$d_{\mathrm{AG}}(C_{\Gamma^{(\mathrm{prec}(s))}}) > 2 \tau$,
then check whether $\alpha_0/\alpha_1 \in \RR$, $\ev(-\alpha_0/\alpha_1) \in C_{\Gamma^{(\mathrm{prec}(s))}}$ and  $$\wt \left(\ev(-\alpha_0/\alpha_1+\sum_{s\leq s' \in \Gamma} w_{s'} \varphi_{s'})- \vec{r}\right) \leq  \tau .$$
If  the previous statement holds, include $\ev(-\alpha_0/\alpha_1+\sum_{s\leq s' \in \Gamma} w_{s'} \varphi_{s'})$ into the list of codewords, and avoid proceeding with
$\mathrm{prec}(s)$. Otherwise, iterate the procedure with $\mathrm{prec}(s)$.

The  procedure above is based on the following observations: 
\begin{itemize}
\item If there exists a codeword $\vec{c} \in C_{\Gamma^{(\mathrm{prec}(s))}}$
with Hamming distance  $\leq \tau$ from $\vec{r}^{(\mathrm{prec}(s))}$,
then, by Proposition \ref{prop10}, executing the iteration on $I_{\vec{r}^{(\mathrm{prec}(s))}}$ gives the only codeword $\vec{c}$ as the list of codewords,
corresponding to $-\alpha_0/\alpha_1$. Therefore, iterations with lower $s$  are meaningless.

\item It was proved in \cite[Lemmas 2.3 and 2.4]{beelen08},  that if $2 \wt (\ev(\beta)- \vec{r}^{(\mathrm{prec}(s))}) + 2g < n-s$ then $\beta$ must appear as $-\alpha_0 / \alpha_1$. Then we can terminate the algorithm at latest $s= \max\{ s \mid 2 \tau + 2g < n-s\}$. Because, under this assumption, any other codeword $\ev(\beta')\in C_{\Gamma^{(\mathrm{prec}(s))}}$
gives $-\alpha'_0/\alpha'_1$ with $-v_Q(\alpha'_1) > -v_Q(\alpha_1)$,
hence $\beta'$ cannot correspond to $f_{\mathrm{min}}$. Note that the genus zero case was proved in \cite[Theorem 12]{kuijper11}. 
\end{itemize}

\item \textit{Termination:} After reaching $s= \max\{ s \mid 2 \tau + 2g < n-s\}$ or after verifying the earlier termination condition, include  the recovered message $(w_{s_1},w_{s_2},\dots,w_{s_k})$ in the output list. 

\end{enumerate}

\subsection{Relation of $\nu(s)$ to \cite{andersen08}}

In \cite{lee11}, $\nu(s)$ was introduced in the same way as in Eq.\ (\ref{eq:nu}). We claim that $\nu (s)$ is equivalent to the sets used in \cite{andersen08,geil11} for bounding the minimum distance. Let $\Gamma_{\mathrm{indep}}$ as in Eq.\   (\ref{eq:improved}).
Let $S_{\mathrm{indep}}=\{ u \mid C_u \neq C_{u-1} \}$. 
Define 
\begin{equation}\label{eq:lambda}\lambda(s) = | \{ j \in S \mid j+s \in S_{\mathrm{indep}} \}|.\end{equation}
The bound in \cite[Propositions 27 and 28]{andersen08} for the minimum distance of $C_{\Gamma}$ is
$$d_{\mathrm{AG}} (C_\Gamma) = \min \{ \lambda(s) \mid s \in \Gamma\} \ge n - s_k.$$

The following proposition implies that $d_u=\min\{\nu(s)\mid s\in S,s\le u\}$  is equivalent to $d_{\mathrm{AG}} (C_u)$, and therefore \cite[Theorem 8]{andersen08} implies \cite[Proposition 12]{lee11}.

\begin{proposition}\label{prop:AG}
Let $s \in S$, one has that $\nu (s) = \lambda (s)$.
\end{proposition}

\subsection{Proof and error correction capability of the algorithm}
We will prove in this section the correctness and error correction capability of the algorithm. Using \cite[Lemmas 6,7 and Proposition 8]{lee11} we have the following proposition that is an extension of \cite[Proposition 10]{lee11}.

\begin{proposition}\label{prop10}
Let $\lambda (s) = \nu(s)$ as in Eqs.\  (\ref{eq:nu}) and (\ref{eq:lambda}). We have
$$	\sum_{w_i=\gw_s}\bar{c}_i\geq \sum_{w_i\neq\gw_s}\bar{c}_i - 2\wt(\vec{e}(\vec{c}))+\lambda (s). $$
\end{proposition}

One has that the set $B^{(s)}$ is a Gr\"obner basis of the $\mathbf{F}_q[x_1]$-module $I_{\vec{r}^{(s)}}$ with respect to $>_s$ by \cite[Proposition 11]{lee11} and combining this with Proposition \ref{prop10},  we obtain the error correction capability of the algorithm in Sec.\  \ref{subsec:Algo} as a unique decoding algorithm. Moreover, it a list-decoding algorithm with error bound $\tau$  by Eq.\  (\ref{eq:acceptedvote}).

\begin{theorem}
Let $\vec{r}=\vec{c}+\vec{e}(\vec{c})$. If $wt (\vec{e}(\vec{c})) \le \tau$ then $\vec{c}$ is in the output list of the algorithm in Sec.\  \ref{subsec:Algo}. If $2 \wt (\vec{e}(\vec{c})) <  d_{\mathrm{AG}} (C_\Gamma )$ then  $w_s = \omega_s$ for all $s \in \Gamma$ and $$\sum_{s \in \Gamma} w_s \varphi_s = \mu,$$
where $\mu$ and $\omega_s$'s are as defined at Eq.\ (\ref{eq:mus}).
\end{theorem}

\subsection{Computer experiments: Comparison against Guruswami-Sudan algorithm}\label{sec34}
We implemented the proposed list decoding algorithm
on Singular \cite{singular} and decoded 1,000
randomly generated codewords with the following conditions.
Firstly we used the one-point primal code $C_u$
with $u=20$ on the Klein quartic
over $\mathbf{F}_8$. It is $[23,18]$ code and its AG bound
\cite{andersen08} is $4$ while
Goppa bound is $3$.
Guruswami-Sudan decoding can decode up to $1$. Our algorithm can list all the codewords within Hamming
distance $2$.
The errors were uniformly randomly generated among the vectors
with Hamming weight $2$ and executed the
decoding algorithm with $\tau=2$. With $757$ transmissions the list size was $1$,
with $180$ transmissions the list size was $2$, and with $63$ transmissions
the list size was $3$, where the list size means the number of codewords
whose Hamming distance from the received word is $\leq \tau$.
The maximum number of iterations was 266, the minimum was 11, the
average was $195.7$, and the standard deviation was $60.5$.

Secondly we used the improved code construction \cite{feng95}
with the designed minimum distance $6$. It is a $[64, 55]$ code.
In order to have the same dimension by $C_u$ we have to
set $u=60$, whose AG bound \cite{andersen08} is $4$ and the Guruwsami-Sudan
can correct 2 errors.
The proposed algorithm finds all codewords in the improved
code with 3 errors.
The errors were uniformly randomly generated among the vectors
with Hamming weight $3$. With $998$ transmissions the list size was $1$,
and with $2$ transmissions the list size was $2$.
The maximum number of iterations was $1128$, the minimum was $14$, the
average was $794.2$, and the standard deviation was $179.8$.

Thirdly we used the same code as the second experiment,
while the errors with Hamming weight $3$
were randomly generated toward another nearest
codeword.
With $901$ transmissions the list size was $2$,
and with $99$ transmissions the list size was $5$.
The maximum number of iterations was $818$, the minimum was $196$, the
average was $754.5$, and the standard deviation was $185.3$.
Observe that the list size cannot become $1$ under this condition,
and the simulation confirmed it.

\section{Conclusion}\label{sec5}
We generalized the two decoding algorithms \cite{lee09,lee11}
to all algebraic curves.
We also extend the latter algorithm \cite{lee11} to a list decoding
one. The resulted list decoding algorithm can correct more errors than
the Guruswami and Sudan algorithm \cite{guruswami99}.
The detailed analysis of the computational complexity of the
latter one is a future research agenda.

\section*{Acknowledgments}
The authors would like to thank an anonymous reviewer for pointing
out errors in the initial manuscript.
This research was partially supported by the MEXT Grant-in-Aid for Scientific Research (A) No.\ 23246071, the Villum Foundation through their VELUX Visiting Professor Programme 2011--2012,  the Danish National Research Foundation and the National Science Foundation of China (Grant No.11061130539) for the Danish-Chinese Center for Applications of Algebraic Geometry in Coding Theory and Cryptography and by Spanish grant MTM2007-64704.


\begin{thebibliography}{10}
\providecommand{\url}[1]{#1}
\csname url@samestyle\endcsname
\providecommand{\newblock}{\relax}
\providecommand{\bibinfo}[2]{#2}
\providecommand{\BIBentrySTDinterwordspacing}{\spaceskip=0pt\relax}
\providecommand{\BIBentryALTinterwordstretchfactor}{4}
\providecommand{\BIBentryALTinterwordspacing}{\spaceskip=\fontdimen2\font plus
\BIBentryALTinterwordstretchfactor\fontdimen3\font minus
  \fontdimen4\font\relax}
\providecommand{\BIBforeignlanguage}[2]{{%
\expandafter\ifx\csname l@#1\endcsname\relax
\typeout{** WARNING: IEEEtranS.bst: No hyphenation pattern has been}%
\typeout{** loaded for the language `#1'. Using the pattern for}%
\typeout{** the default language instead.}%
\else
\language=\csname l@#1\endcsname
\fi
#2}}
\providecommand{\BIBdecl}{\relax}
\BIBdecl

\bibitem{adams94}
W.~W. Adams and P.~Loustaunau, \emph{An Introduction to Gr\"obner Bases},
  ser. Graduate Studies in Mathematics.\hskip 1em plus 0.5em minus 0.4em\relax
  Providence, RI: American Mathematical Society, 1994, vol.~3.

\bibitem{kuijper11}
M.~Ali and M.~Kuijper, ``A parametric approach to list decoding of
  {Reed-Solomon} codes using interpolation,'' \emph{IEEE Trans.\ Inform.\
  Theory}, vol.~57, no.~10, pp. 6718--6728, Oct. 2011, arXiv:1011.1040.

\bibitem{andersen08}
H.~E. Andersen and O.~Geil, ``Evaluation codes from order domain theory,''
  \emph{Finite Fields Appl.}, vol.~14, no.~1, pp. 92--123, Jan. 2008.

\bibitem{beelen10}
P.~Beelen and K.~Brander, ``Efficient list decoding of a class of
  algebraic-geometry codes,'' \emph{Adv. Math. Commun.},
  vol.~4, no.~4, pp. 485--518, 2010.

\bibitem{beelen08}
P.~Beelen and T.~H\o holdt, ``The decoding of algebraic geometry codes,'' in
  \emph{Advances in Algebraic Geometry Codes}, ser. Coding Theory and
  Cryptology, E.~Mart\'inez-Moro, C.~Munuera, and D.~Ruano, Eds.\hskip 1em plus
  0.5em minus 0.4em\relax World Scientific, 2008, vol.~5, pp. 49--98.

\bibitem{feng95}
G.~L. Feng and T.~R.~N. Rao, ``Improved geometric Goppa codes part I, basic
  theory,'' \emph{IEEE Trans. Inform. Theory}, vol.~41, no.~6, pp. 1678--1693,
  Nov. 1995.

\bibitem{gmr12a}
O.~Geil, R.~Matsumoto, and D.~Ruano,
``List decoding algorithm based on voting in Gr\"obner bases for
general one-point AG codes,'' arXiv:1203.6127, Apr. 2012.

\bibitem{gmr12b}
------,
``Generalization of the Lee-O'Sullivan list decoding for one-point AG codes,'' arXiv:1203.6129, Apr. 2012.

\bibitem{geil11}
O.~Geil, C.~Munuera, D.~Ruano, and F.~Torres, ``On the order bounds for
  one-point {AG} codes,'' \emph{Adv. Math. Commun.}, vol.~5, no.~3, pp.
  489--504, 2011.

\bibitem{geilpellikaan00}
O.~Geil and R.~Pellikaan, ``On the structure of order domains,'' \emph{Finite
  Fields Appl.}, vol.~8, no.~3, pp. 369--396, Jul. 2002.

\bibitem{singular}
\BIBentryALTinterwordspacing
G.-M. Greuel, G.~Pfister, and H.~Sch\"onemann, ``{\sc Singular} 3.0,'' Centre
  for Computer Algebra, University of Kaiserslautern, {A Computer Algebra
  System for Polynomial Computations}, 2005. [Online]. Available:
  \url{http://www.singular.uni-kl.de}
\BIBentrySTDinterwordspacing

\bibitem{guruswami99}
V.~Guruswami and M.~Sudan, ``Improved decoding of Reed-Solomon and
  algebraic-geometry codes,'' \emph{IEEE Trans.\ Inform.\ Theory}, vol.~45,
  no.~4, pp. 1757--1767, Sep. 1999.


\bibitem{lee12}
K.~Lee, ``Unique decoding of plane AG codes revisited,''
arXiv:1204.0052, Mar. 2012.

\bibitem{lee11}
\BIBentryALTinterwordspacing
K.~Lee, M.~Bras-Amor\'os, and M.~E. O'Sullivan, ``Unique decoding of plane AG
  codes via interpolation,'' 2012, IEEE Trans. Inform. Theory, Early Access.
  [Online]. Available: \url{http://dx.doi.org/10.1109/TIT.2012.2182757},
  arXiv:1110.6251.
\BIBentrySTDinterwordspacing

\bibitem{lee09}
K.~Lee and M.~E. O'Sullivan, ``List decoding of Hermitian codes using
  Gr\"obner bases,'' \emph{J.\ Symbolic Comput.}, vol.~44, no.~12, pp.
  1662--1675, Dec. 2009, arXiv:cs/0610132.

\bibitem{little11}
J.~B. Little, ``List decoding for AG codes using Gr\"obner bases,''
  presented at
  \emph{SIAM Conference on Applied Algebraic Geometry}, North Carolina State
  University, NC, USA, Oct. 2011.

\bibitem{matsumoto99ldpaper}
R.~Matsumoto and S.~Miura, ``Finding a basis of a linear system with pairwise
  distinct discrete valuations on an algebraic curve,'' \emph{J.\ Symbolic
  Comput.}, vol.~30, no.~3, pp. 309--323, Sep. 2000.

\bibitem{miuraform}
\BIBentryALTinterwordspacing
------, ``On construction and generalization of algebraic geometry codes,'' in
  \emph{Proc.\ Algebraic Geometry, Number Theory, Coding Theory, and
  Cryptography}, T.~Katsura \emph{et~al.}, Eds., Univ. Tokyo, Japan, Jan. 2000,
  pp. 3--15. [Online]. Available:
  \url{http://www.rmatsumoto.org/repository/weight-construct.pdf}
\BIBentrySTDinterwordspacing

\bibitem{miura92}
S.~Miura, ``Algebraic geometric codes on certain plane curves,''
  \emph{Electronics and Communications in Japan (Part III: Fundamental
  Electronic Science)}, vol.~76, no.~12, pp. 1--13, Dec. 1993.

\bibitem{miura98}
------, ``Linear codes on affine algebraic
  curves,'' \emph{Trans. IEICE}, vol. J81-A,
  no.~10, pp. 1398--1421, Oct. 1998 (Japanese).

\bibitem{saints95}
K.~Saints and C.~Heegard, ``Algebraic-geometric codes and multidimensional
  cyclic codes: A unified theory and algorithms for decoding using Gr\"obner
  bases,'' \emph{IEEE Trans. Inform. Theory}, vol.~41, no.~6, pp. 1733--1751,
  Nov. 1995.

\bibitem{tang98}
L.-Z. Tang, ``A Gr\"obner basis criterion for birational equivalence of affine
varieties,'' \emph{J. Pure Appl. Algebra}, vol.~123, pp.~275--283, Jan. 1998.
\end{thebibliography}


\end{document}